\documentstyle[preprint,aps]{revtex}

\input{epsf}

\def\be{\begin{equation}}
\def\ee{\end{equation}}
\def\ba{\begin{eqnarray}}
\def\ea{\end{eqnarray}}

\def\ltsima{$\; \buildrel < \over \sim \;$}
\def\simlt{\lower.5ex\hbox{\ltsima}}
\def\gtsima{$\; \buildrel > \over \sim \;$}
\def\simgt{\lower.5ex\hbox{\gtsima}}

\begin{document}
\title{{\bf Radiation-condensation instability in a highly ionized dusty plasma}}
\author{Miguel H. Ib\'{a}\~{n}ez S.\thanks{%
ibanez@ciens.ula.ve}}
\address{{\it Centro de Astrof\'{\i}sica Te\'{o}rica, Facultad de Ciencias,}\\
{\it Universidad de los Andes, Apartado 26, Ipostel, La Hechicera, M\'{e}rida%
}, {\it Venezuela}\\
and Yuri A. Shchekinov \\
{\it Department of Physics, University of Rostov,Rostov on Don, 34409, }%
Russia}
\maketitle

\begin{abstract}
The dynamics of linear perturbations in a radiatively cooling dusty plasma
is considered, with the charge of both dust ($Z_d$) and plasma ($Z_p$)
components being allowed to vary with their densities. It is shown that in
the long-wavelength limit corresponding to the characteristic cooling
length, when the plasma can be treated as quasineutral, the presence of dust
particles changes the criteria for radiation instability, regardless the
charging process of the dust particles. In particular, the condensation
(isobaric) mode is shown to be stabilized (destabilized) if in the
equilibrium, the relation between densities of the dust $n_d$ and plasma $n$
under the quasineutrality condition, $(d\ln n_{d}/d \ln n)_{q}<1$ ($>1$) is
satisfied, while the isentropic mode is stabilized (destabilized) when the
opposite inequalities take place; the isochoric mode is unaltered. Numerical
estimates show that these effects can be important in hot phases ($T\sim
10^{6}$ K) of the interstellar plasma, and in tokamak plasma near the walls.


%
%
%

\end{abstract}

\section*{I. INTRODUCTION}

It is well known (see, for example, Refs. 1, 2, 3) that the presence of
charged dust particles in plasma changes its dispersion properties
qualitatively, and that the dust itself can give rise to new plasma
instabilities. From intuitive point of view it seems possible that dust must
be important in dynamics of radiatively cooling plasmas, not only because it
absorbs radiation, but also due to its ability to transform thermal energy
of the plasma into radiation in inelastic collisions with electrons and
ions. Apparently, it can be of great importance in high temperature ($T\lower%
.5ex\hbox{\gtsima}10^{6}$ $K$) cosmic and laboratory (in particular, close
to tokamak walls) plasmas, where frequent collisions with the ions maintain
dust particles sufficiently hot, so that they re-emit thermal energy
efficiently$^{4}$. The present work is aimed to analyze the effects of
charged dust particles on the conditions for the radiation instabilities. It
is important to note that this issue has been addressed recently in$^{5,6}$,
however in these papers dust particles were considered only from the point
of view of electrostatic interactions, while the cooling processes were
treated as independent on the presence of dust. In our study we explicitly
assume that dust particles contribute substantially to the net cooling rate,
which brings to the system qualitatively new dynamical features. For the
sake of simplicity we will consider here only highly ionized plasmas, when
the effects from neutral particles can be neglected. In Sec.~II.A we briefly
discuss how important can be dust as a cooler of the plasma. In Sec. II.B we
formulate a simplified two-fluid model of a collisional and radiatively
cooling dusty plasma. In Sec. III.A we describe the steady state
equilibrium, with the relevant parameters estimated in Sec. III.B, while in
Sec. III.C linearized equations are written. In Sec. III.D the dispersion
equation and its solutions are given and criteria of thermal instability in
different wavelength ranges are discussed. Sec. IV summarizes the results.

\section*{II. BASIC EQUATIONS}

\subsection*{A. Radiatively cooling dusty plasma}

\noindent In an optically thin plasma dust particles, if present, besides
their dynamical effects through electrostatic interaction with plasma
particles, can also work as a cooling agent via the collisional transfer of
thermal energy of plasma component into infrared radiation. If one assumes,
as an idealization, that dust particles emit as black body, the energy loss
rate due to dust particles varies as $\propto T_{d}^{4}$, where $T_{d}$ is
temperature of the dust. For realistic particles this dependence can be
steeper. Detailed calculations of the relevant cooling rate are made in Ref.
4. They show that for physical conditions and abundance of dust typical for
the interstellar medium (ISM), radiation losses from dust particles
dominates cooling from the line emission of heavy elements by two orders of
magnitude in the interval $T=10^{6}-10^{10}$ $K^{4}$. One should stress,
however, that at temperatures $T\sim 3\times (10^{7}-10^{8})$ $K$ dust
particles are efficiently destroyed in collisions with the ions and electrons%
$^{7}$, and therefore only in temperature range between $10^{6}$ $K$ and $%
3\times 10^{7}$ $K$ dust contributes sufficiently to the net cooling.
Apparently, dust cooling can also be important in tokamak plasma near the
walls, where temperature can lie in this interval, or in regions of MARFEs
(multifaceted asymmetric radiation from the edge) formation if radiating
impurities are introduced into the plasma by sputtering.

\subsection*{B. Two fluids}

We will consider a simplified two-fluid approach, with the dust charge
instantaneously (on times much shorter than other characteristic time scales
of the system) settled on the equilibrium $Z_d=Z_d(n,n_d,T)$, where $n$ and $%
n_d$ are the plasma and dust densities, $T$ is the plasma temperature. To
outline the main physical effects, we will study the simple case of
potential perturbations in the planar geometry. The full set of equations in
this case is

\begin{equation}
\partial_t n+\partial_x (nv)=0,
\end{equation}

\begin{equation}
\rho\partial_t v+\rho v\partial_x v=-\partial_x p -eZ_pn\partial_x
\phi-f_0(v-u),
\end{equation}

\begin{equation}
{\frac{3}{2}}d_t p-{\frac{5}{2}}{\frac{p}{n}}d_tn=-{\cal L}(T,n,n_d),
\end{equation}

\begin{equation}
\partial_t n_d+\partial_x (n_du)=0,
\end{equation}

\begin{equation}  \label{euldust}
\rho_d\partial_t u+\rho_d u\partial_x u= -eZ_dn_d\partial_x \phi+f_0(v-u),
\end{equation}

\begin{equation}
p=k_BnT,
\end{equation}

\begin{equation}
\partial _{x}^{2}\phi =-4\pi e(Z_{d}n_{d}+Z_{p}n),
\end{equation}
where $v$, $u$ are velocities of the plasma and dust particles,
respectively, $f_{0}$ [$g$ $cm^{-3}s^{-1}$] is the coefficient describing
the friction between dust and plasma, $\rho =mn$ and $\rho _{d}=m_{d}n_{d}$
are the mass densities of the plasma and dust components, $p$, the gas
pressure, $\phi $, the electrostatic potential, $Z_{d}$, the dust charge
(negative for negatively charged dust), $Z_{p}$, the mean plasma charge
defined as $Z_{p}n=(Z_{i}n-n_{e})=(Z_{i}-x)n$ with $Z_{i}$ being the ion
charge, and $x=n_{e}/n$; note that for positively charged dust particles $%
x\geq Z_{i}$; here an adiabatic index $\gamma =5/3$ has been assumed and the
net cooling function ${\cal L}$ explicitly depending on number density of
dust particles; the dust component is assumed kinetically cold, so that the
dust pressure term in Eq. \ref{euldust} is omitted; $\partial _{a}$ and $d_{a}$
denote partial and Lagrangian derivatives with respect to $a$. In general,
dust particles contribute both to heating ({\it e.g.~} via photoemission of
electrons from grain surface, see for recent discussion Ref. 8), and cooling
(via collisional transfer of kinetic energy of plasma particles into heat of
dust grains and subsequent re-emission in, as a rule, infrared range$^{4}$).
It is obvious that the charge separation are important only on the spatial
scales comparable (or shorter) to Debye length, on longer lengths plasma can
be treated as quasineutral (see, below).

\section*{III. DISPERSION RELATION}

\subsection*{A. Unperturbed state}

We assume that in an unperturbed state radiative cooling is balanced by
energy input from external sources ({\it e.g.~} hard emission, Ohmic heating
etc.), so that

\begin{equation}
{\cal L}=0,
\end{equation}
[note that in most cases the net cooling function can be presented in the
form ${\cal L}=L(T,n,n_d)n-\Gamma(T,n,n_d)n$, where $L(T,n,n_d)n$ is the
radiative cooling rate, $\Gamma(T,n,n_d)n$, the heating rate due to external
energy sources]. We assume also that unperturbed plasma is quasineutral

\begin{equation}  \label{eqcharge}
Z_{d0}n_{d0}+Z_{p0}n_0=0.
\end{equation}
We will further assume that the dust and plasma charges adjust the
corresponding equilibria instantaneously $Z_d=Z_d(n,n_d,T)$ and $%
Z_p=Z_p(n,n_d,T)$, {\it ~i.e.~} that the charging characteristic times are
shorter than other relevant times.

\subsection*{B. Evaluation of the relevant numbers}

The problem is characterized by three time scales: thermal, frictional, and
electrostatic. Thermal time scale is written as

\begin{equation}
\tau_0={\frac{k_BT_0}{L_0}},
\end{equation}
here $L_0$ is the radiative cooling rate taken at the equilibrium, $k_B$,
Boltzmann constant. Frictional times for the plasma and dust particles

\begin{equation}
\tau_f={\frac{\rho_0}{f_0}}, \quad \tau_{f,d}={\frac{\rho_{d0}}{f_0}},
\end{equation}
respectively, and electrostatic time (the inverse plasma frequency)

\begin{equation}
\tau_e=\sqrt{\frac{m}{4\pi e^2n}}.
\end{equation}
All characteristic times (frictional and electrostatic) relevant to the dust
component are connected with the corresponding plasma times by the intrinsic
constants $\delta=n_{d0}/n_0$, the concentration of dust particles, and $%
\mu=m_dn_{d0}/mn_0\equiv \rho_{d0}/\rho_{0}$, the mass density ratio of the
dust and plasma components:

\begin{equation}
\tau _{{f,d}}=\mu \tau _{f},\quad \tau _{e,d}=\delta ^{-1}\mu ^{-1/2}\tau
_{e}.
\end{equation}

It is readily seen that two dimensionless parameters appear in the problem

\begin{equation}
\kappa={\frac{\tau_0}{\tau_e}}={\frac{k_BT_0}{L_0}}\sqrt{\frac{4\pi e^2n}{m}}%
, ~{\rm and}~\nu={\frac{\tau_0}{\tau_f}}={\frac{k_BT_0f_0}{L_0\rho_0}},
\end{equation}
which characterize the role of electrostatic and frictional forces in
dynamics of radiation-condensation instability.

\subsection*{C. Linearized equations}

The full set of linearized dynamical equations for a nonmagnetized dusty
plasma is written as

\begin{equation}  \label{plasmacon}
\partial_tn+n_0\partial_xv=0,
\end{equation}

\begin{equation}  \label{plasmamot}
mn_0\partial_tv+\partial_xp+eZ_{p0}n_0\partial_x\phi+f_0(v-u)=0,
\end{equation}

\begin{equation}  \label{plasmaen}
{\frac{3}{2}}d_tp-{\frac{5}{2}}{\frac{p_0}{n_0}}d_tn+(\partial_n {\cal L})n
+(\partial_{n_d}{\cal L})n_d+(\partial_T{\cal L})T=0,
\end{equation}

\begin{equation}  \label{dustcon}
\partial_tn_d+n_{d0}\partial_xu=0,
\end{equation}

\begin{equation}  \label{dustmot}
m_dn_{d0}\partial_tu+eZ_{d0}n_{d0}\partial_x\phi-f_0(v-u)=0,
\end{equation}

\begin{equation}  \label{eqstate}
{\frac{p}{p_0}}={\frac{n}{n_0}}+{\frac{T}{T_0}},
\end{equation}

\begin{eqnarray}  \label{poisson}
\partial _{x}^{2}\phi &=&-4\pi e[Z_{d0}n_{d}+(\partial
_{n}Z_{d})n_{d0}n+(\partial _{T}Z_{d})n_{d0}T+(\partial
_{n_{d}}Z_{d})n_{d0}n_{d}+  \nonumber \\
&&Z_{p0}n+(\partial _{n}Z_{p})n_{0}n+(\partial _{T}Z_{p})n_{0}T+(\partial
_{n_{d}}Z_{p})n_{0}n_{d}],
\end{eqnarray}
where the unperturbed variables are given with subscript zero, $a_{0}$,
while the perturbations are without subscript, $a$.

In order to evaluate the interrelation between electrostatic and radiative
effects it is useful to write the linearized Poisson equation \ref{poisson}
 in a non-dimensional form

\begin{eqnarray}
\partial _{\xi }^{2}\bar{\phi} &=&-\kappa ^{2}[\delta \{Z_{d0}\bar{n}%
_{d}+(\partial _{n}Z_{d})\bar{n}+(\partial _{n_{d}}Z_{d})\bar{n}%
_{d}+(\partial _{T}Z_{d})\bar{T}\}+  \nonumber \\
&&(\partial _{n}Z_{p})\bar{n}+(\partial _{n_{d}}Z_{p})\bar{n}_{d}+(\partial
_{T}Z_{p})\bar{T}+Z_{p0}\bar{n}],
\end{eqnarray}
here all variables with bar are normalized to unperturbed values: $\bar{a}%
=a/a_{0}$; $\tau =t/\tau _{0}$, $\xi =x/\tau _{0}c_{0}$, where $c_{0}$ is
the isothermal sound speed, the potential is normalized to $mc_{0}^{2}/e$, $%
\kappa $ is defined above.

For typical interstellar or tokamak hot plasma at $T\sim 10^{6}$ $K$, $%
\kappa $ can be estimated as

\begin{equation}
\kappa ^{2}\sim {\frac{3\cdot 10^{30}}{n_{0}}},
\end{equation}
It is readily seen from these estimates that motions on electrostatic and
radiation scales can be separated due to a huge difference in corresponding
times and lengths. This simply means that electrostatic forces acting
between the charged dust and plasma components keep dusty plasma
quasineutral on much shorter time scales than the relevant hydrodynamical
scales. Formally it can be expressed in the form of perturbation procedure
applied to the above dimensionless linearized equations with dynamical
variables as a series

\begin{equation}
a=a^{0}+\kappa^{-2}a^{1}+..,
\end{equation}
and potential as

\begin{equation}
\phi =\kappa ^{-2}\phi ^{1}+\kappa ^{-4}\phi ^{2}+...
\end{equation}
Thus in what follows we turn to the dimensional variables and equations,
with the quasineutrality condition instead of solving Poisson equation.

\subsection*{D. Dispersion equation}

In this framework the quasineutrality equation can be written in terms of $%
n_{d}$

\begin{eqnarray}  \label{vard}
n_{d} &=&-[Z_{d0}+(\partial _{n_{d}}Z_{d})n_{d0}+(\partial
_{n_{d}}Z_{p})n_{0}]^{-1}[(\partial _{n}Z_{d})n_{d0}+Z_{p0}+(\partial
_{n}Z_{p})n_{0}]n+  \nonumber \\
&&-[Z_{d0}+(\partial _{n_{d}}Z_{d})n_{d0}+(\partial
_{n_{d}}Z_{p})n_{0}]^{-1}[(\partial _{T}Z_{d})n_{d0}+(\partial
_{T}Z_{p})n_{0}]T.
\end{eqnarray}
From Eq. (\ref{vard}) it follows that $(\partial
_{T}n_{d})_{q}=-[Z_{d0}+(\partial _{n_{d}}Z_{d})n_{d0}+(\partial
_{n_{d}}Z_{p})n_{0}]^{-1} [(\partial _{T}Z_{d})n_{d0}+(\partial
_{T}Z_{p})n_{0}]$, which will be taken as zero in order that the
quasineutrality to hold for purely isochoric perturbations; here the
subscript $q$ denotes that the derivative is taken over the quasineutrality
state.

Combining Equations of motion \ref{plasmamot} and \ref{dustmot} in order
to eliminate undefined $\phi $ we arrive finally for perturbations in the
form $a\propto \exp (\Omega t+ikx)$ \ to the eigenmatrix ${\bf M}$ for the
eigenvector $(n^{\prime },T^{\prime },v^{\prime },u^{\prime })$ 
\begin{equation}
{\bf M=}\left( 
\begin{array}{cccc}
\Omega & 0 & ikn_{0} & 0 \\ 
ik\frac{p_{0}}{n_{0}} & ik\frac{p_{0}}{T_{0}} & \rho _{0}\Omega & \rho
_{d0}\Omega \\ 
-\frac{p_{0}}{n_{0}}\Omega +{\cal L}_{n}+A{\cal L}_{n_{d}} & \frac{3}{2}%
\frac{p_{0}}{T_{0}}\Omega +{\cal L}_{T} & 0 & 0 \\ 
A\Omega & 0 & 0 & ikn_{d0}
\end{array}
\right) ~,  \label{Matrix}
\end{equation}
where $A=-[Z_{d0}+(\partial _{n_{d}}Z_{d})n_{d0}+(\partial
_{n_{d}}Z_{p})n_{0}]^{-1}[(\partial _{n}Z_{d})n_{d0}+Z_{p0}+(\partial
_{n}Z_{p})n_{0}]\equiv (dn_{d}/dn)_{q}$. Finally the compatibility condition 
$\det ({\bf M})=0$ leaves the dispersion equation

\begin{equation}
\Omega ^{3}+{\frac{2}{3}}{\frac{T_{0}}{p_{0}}}{\cal L}_{T}\Omega ^{2}+{\frac{%
5}{3}}k^{2}c_{d}^{2}\Omega +{\frac{2}{3}}k^{2}c_{d}^{2}\Biggl[{\frac{T_{0}}{%
p_{0}}}{\cal L}_{T}-{\frac{n_{0}}{p_{0}}}\left( {\cal L}_{n}+A{\cal L}%
_{n_{d}}\right) \Biggr]=0,  \label{dispersion}
\end{equation}
where $c_{d}=c_{0}/\sqrt{1+\mu }$, $c_{0}$ being the isothermal sound speed
in a plasma gas. It is seen that this equation is identical to the standard
dispersion equation$^{9}$, when all terms correspondent to dust component
are omitted.

\subsubsection*{D1. Isobaric mode}

\noindent In general, there are three solutions of Eq. \ref{dispersion}.
One corresponds to the so-called isobaric (condensation) mode, which
formally can be obtained from Eq. \ref{dispersion} putting $k\rightarrow
\infty $ and $Im(\Omega )=0$ (see discussion in Refs. 9, 10), i.e.

\begin{equation}  \label{growth}
\Omega\sim {\frac{2}{5p_0}}[n_0({\cal L}_n+A{\cal L}_{n_d})-T_0{\cal L}_T].
\end{equation}
The condition for instability of this mode is also straightforwardly follows
from the Hurwitz criterion that the last coefficient of the third order
polynomial in Eq. \ref{dispersion} is negative, i.e. $n_0({\cal L}_n+A{\cal L}%
_{n_d})- T_0{\cal L}_T>0$.

For the cooling function in the form ${\cal L}=\Lambda(T)n^2+
\Lambda_d(T)nn_d-\Gamma n$, one can write the growth rate as

\begin{equation}  \label{growc}
\Omega\sim {\frac{2{\cal L}^c}{5p_0}}\Biggl\{\Biggl[1+ \biggl[\left({\frac{d
\ln n_d}{d \ln n}}\right)_q-1\biggr] \eta_d\Biggr]-{\frac{\partial \ln {\cal %
L}^c}{\partial \ln T}}\Biggr\},
\end{equation}
where ${\cal L}^c=\Lambda(T)n^2+\Lambda_d(T)nn_d$, $\eta_d=\Lambda_d(T)nn_d/ 
{\cal L}^c$ is the relative contribution of dust cooling. It is readily seen
that for strongly dynamically coupled ions and dust when $n_d=\delta n$,
electrostatic forces do not affect radiation instability. When $(d \ln n_d/d
\ln n)_q<1$, i.e. quasineutrality and dust charging requires that the dust
particles escape compressed regions, the condensation mode is stabilized
because the corresponding condition

\begin{equation}
{\frac{\partial \ln {\cal L}^{c}}{\partial \ln T}}<1-\Biggl|{\frac{d \ln
n_{d}}{d \ln n}}-1\Biggr|\eta _{d},
\end{equation}
fulfills in a narrower temperature interval than when electrostatic effects
disappear. Contrary, if quasineutrality and dust charging requires an
enhancement of dust particles in compressed regions, i.e. $(d \ln n_{d}/d
\ln n)_{q}>1$, the condensation instability is enhanced. Qualitatively it
can be understood as due to an enhancement of the net cooling rate
associated with dust cooling.

\subsubsection*{D2. Isentropic (acoustic) mode}

\noindent The other two modes in the short-wavelength limit correspond to
overstable acoustic motions with the wave numbers $\pm k$ and the growth
rate much smaller than the wave frequency: $|Re(\Omega)|\ll
|Im(\Omega)|^{9,10}$. In order to find the solution in this case we will
expand it as

\begin{equation}
\Omega=\pm i\left({\frac{5}{3}}\right)^{1/2}kc_d+\Omega_r,
\end{equation}
where $\Omega_r$ is the growth rate of the amplitude of acoustic waves, for
which we obtain to the first order to $k$

\begin{equation}
\Omega_r\sim -{\frac{2}{15p_0}}\Biggl[T_0{\cal L}_T+{\frac{3}{2}} n_0({\cal L%
}_{n}+A{\cal L}_{n_d})\Biggr],
\end{equation}
or 
\begin{equation}  \label{growad}
\Omega_r\sim -{\frac{2{\cal L}^c}{15p_0}}\Biggl\{ {\frac{\partial \ln {\cal L%
}^c}{\partial\ln T}}+{\frac{3}{2}}\Biggl[1+ \biggl[\left({\frac{d\ln n_d}{%
d\ln n}}\right)_q-1\biggr]\eta_d\Biggr]\Biggr\}.
\end{equation}
It is seen from (\ref{growad}) that contrary to the isobaric mode,
isentropic perturbations are destabilized if $(d\ln n_d/d\ln n)_q<1$, since
the restrictions on the cooling rate ${\cal L}^c$ is here weaker than in the
absence of electrostatic forces. Instead, isentropic perturbations are
stabilized when $(d\ln n_d/d\ln n)_q>1$. Such a behavior is qualitatively
clear because the instability of isentropic mode is physically connected
with overheating of adiabatically compressed regions, and as soon as dust
cooling decreases in compressed regions when $(d\ln n_d/d\ln n)_q<1$, it
results in an additional overheating.

\subsubsection*{D3. Isochoric mode}

\noindent The instability criterion for the isochoric mode corresponding to
the long-wavelength limit $k\to 0$: ${\cal L}_T<0$, remains obviously
unaltered by the electrostatic forces and additional dust cooling, because
this limit describes perturbations with $T\neq 0$ and $n_d\simeq n\simeq 0$.

\section*{V. SUMMARY}

\noindent In the present paper we considered radiation instability in an
optically thin dusty plasma with dust particles contributing substantially
to the radiative energy losses. We have shown that combination of the two
effects: electrostatic interaction of the dust and plasma components and
their tight dynamical coupling (expressed in quasineutrality), and
additional energy losses connected with dust particles, strongly modifies
the criteria for the instability. This influence differs for different
instability regimes: the isobaric (condensation) mode is shown to be
destabilized in dusty plasmas if the dust particles are excessively
attracted in the regions of gas compression, $(d \ln n_{d}/d \ln n)_{q}>1$,
and stabilized in the opposite case; in the short-wavelength limit
corresponding to the isentropic (adiabatic) mode effects from the presence
of dust work differently: they quench the instability when $(d \ln n_{d}/d
\ln n)_{q}>1$, and destabilize it in the opposite case; the condition for
the instability remains unaltered in a long-wavelength range where isochoric
mode dominates.

In a simplest case of a dilute hot plasma with dust particles charged
predominantly collisionally, {\it i.e.~} $j_p\to 0$, $(d\ln n_d/d\ln n)_q$
approaches unity from below as (see Appendix)

\begin{equation}
\left({\frac{d\ln n_d}{d\ln n}}\right)_q=1-\left({\frac{m_i}{m_e}}%
\right)^{1/2}j_p,
\end{equation}
and therefore the isobaric mode is quenched, while the isentropic is weakly
enhanced.

As $j_{p}$ approaches the critical value $j_{pc}=(m_{i}/m_{e})^{1/2}-1$,
when dust particles change the charge from negative to positive, $(d\ln
n_{d}/d\ln n)_{q}\rightarrow -\infty $ at $j_{p}=j_{pc}-0$, and $(d\ln
n_{d}/d\ln n)_{q}\rightarrow +\infty $ at $j_{p}=j_{pc}+0$, (see Appendix).
Therefore, from the left side of the critical point, $j_{p}=j_{pc}-0$ where
the charging is dominated by collisions, the isobaric mode is strongly
quenched, while the isentropic is strongly enhanced. Contrary, from the
right side of the critical point, $j_{p}=j_{pc}+0$ where the grains are
charged predominantly by photoionization, the isobaric mode is enhanced,
while the isentropic one is suppressed.

One can therefore conclude that in hot and dilute astrophysical plasmas
embedded in a strong radiation field, where radiation dominates dust
charging, the thermally unstable isobaric mode is enhanced. Under the
conditions of an exposing radiation field with an intensity well below the
critical value (such as in astrophysical plasmas far from strong radiation
sources and dusty plasma sheaths near the tokamak walls) when the
collisional processes are the principal charging mechanisms, isentropic mode
is destabilized, while the isobaric one is quenched.

\section*{\bf Acknowledgments}

We gratefully acknowledge to the anonymous referee  for his critical and
valuable remarks. This work was supported by CDCHT--Universidad de Los
Andes. YS acknowledges partial financial support from the RFBR (project No
99-02-16938) and INTAS (project No 1667) Foundations.

\section*{\bf Appendix}

\noindent Generally, the equation for dust charge is written as ({\it e.g.~}
Refs. 1, 7, 11, 12)

\begin{equation}
j_{p}+(1-y)(1-Qy)=\left( {\frac{m_{i}}{m_{e}}}\right) ^{1/2}(1+Qy)\exp (y), 
\eqnum{A1}  \label{dcharge}
\end{equation}
where $y=e(\phi _{s}-\langle \phi \rangle )/kT$, $\phi _{s}$ is the grain
surface potential, $\langle \phi \rangle $ is the average potential in
plasma, $j_{p}=I_{p}/(ev_{i}a^{2}n)$, the photoionization rate for the
grains in units of the ion charge flux on the surface, $Q=4\pi \lambda
_{D}^{2}n_{d}C$, $C=a(1+a/\lambda _{D})$ is the capacity of a dust grain, $%
\lambda _{D}=\sqrt{kT/[4\pi (n_{e}+n)e^{2}]}$, the Debye length; for a
dilute dusty plasma, $4\pi a^{3}n_{d}/3\ll 1$ and $a\ll \lambda _{D}$, $%
Z_{d}=kTay/e^{2}$ and $C=a$. It is readily seen from Eq. \ref{dcharge}
(see discussion in Refs. 1, 7, 12) that at $j_{p}\rightarrow 0$ dust charge
is negative, $Z_{d}<0$, however, when radiation flux increases (or $n$
decreases) dust grains become charged positively$^{12}$. The dependence $%
n_{d}(n)$ along the quasineutrality condition can be found as the solution
of Eq. \ref{dcharge} along with the quasineutrality condition \ref
{eqcharge} written here as

\begin{equation}
1-x=-{\frac{y}{y_{0}}}\zeta ,  \eqnum{A2}  \label{eqx}
\end{equation}
where for a dilute plasma $y_{0}=e^{2}/(k_{B}Ta)$, and $\zeta =n_{d}/n$.
With these notations $Q=\zeta /[(1+x)y_{0}]$. Equations \ref{dcharge} and 
\ref{eqx} must be solved with a complementary equation for the balance of
electrons

\begin{equation}
a_{r}x=j_{p}\zeta -\left( {\frac{m_{i}}{m_{e}}}\right) ^{1/2}(1+Qy)\exp
(y)+b_{e}x_{a}x+\bar{j}_{p}x_{a},  \eqnum{A3}  \label{elbal}
\end{equation}
which we assume to be dominated by photoemission of electrons from dust
grains, photoionization of neutrals by collisions with thermal electrons,
photoionization of neutrals by external radiation, radiative recombination,
and sticking of electrons on dust grains. Here $a_{r}=\alpha _{r}/(4\pi
v_{i}a^{2})$, $\alpha _{r}$ is radiative recombination coefficient, $%
b_{e}=\beta _{e}/(4\pi v_{i}a^{2})$, $\beta _{e}$ is the rate of collisional
ionizations of neutrals, $\bar{j}_{p}=\bar{I}_{p}/(v_{i}a^{2}n)$ is the
photoionization rate of neutrals, $x_{a}=n_{a}/n$, their fraction.

Assuming that $|1-x|\ll 1$, and taking into account that for hot plasma ($%
T\geq 10^{6}$ $K$) $a_{r},b_{e}\sim 10^{-12}-10^{{-11}}\ll \sqrt{m_{i}/m_{e}}
$, one can reduce Eqs. \ref{dcharge}--\ref{elbal} to the following system

\begin{equation}
{\frac{(1-x)y_{0}}{\zeta }}=\ln {\frac{(m_{i}/m_{e})^{1/2}}{1+j_{p}}}, 
\eqnum{A4}  \label{A4}
\end{equation}

\begin{equation}
x={\frac{j_{p}\zeta +\bar{j}_{p}x_{a}}{(m_{i}/m_{e})^{1/2}\zeta }}, 
\eqnum{A5}  \label{A5}
\end{equation}
which has for $x_{a}\ll 1$ one physically meaningful root

\begin{equation}
\zeta \simeq \lbrack (m_{i}/m_{e})^{1/2}-j_{p}]y_{0}(m_{e}/m_{i})^{1/2}%
\Biggl[\ln {\frac{(m_{i}/m_{e})^{1/2}}{1+j_{p}}}\Biggr]^{-1}.  \eqnum{A6}
\label{A6}
\end{equation}
The other root vanishes at $j_{p}$ and thus is unphysical. It is seen that
at $j_{p}=j_{pc}-1=[(m_{i}/m_{e})^{1/2}-1]-0$ the solution $\zeta
\rightarrow +\infty $ logarithmically, while from the right side $%
j_{p}=j_{pc}+1$ as $\zeta \rightarrow -\infty $. The derivative

\begin{equation}
\left( \frac{dn_{d}}{dn}\right) _{q}=n\left( \frac{d\zeta }{dn}\right)
+\zeta \sim -O\left( [\ln {\frac{(m_{i}/m_{e})^{1/2}}{1+j_{p}}}]^{-2}\right)
+O\left( [\ln {\frac{(m_{i}/m_{e})^{1/2}}{1+j_{p}}}]^{-1}\right) , 
\eqnum{A7}  \label{A7}
\end{equation}
and hence always goes to $-\infty $ at $j_{p}\sim j_{pc}$. Note, that at
this point $x\sim 1-(m_{e}/m_{i})^{1/2}$, so that $|1-x|\ll 1$ holds. The
logarithmic derivative

\begin{equation}
\left( {\frac{n}{n_{d}}}{\frac{dn_{d}}{dn}}\right) _{q}\sim -{\frac{j_{p}}{%
1+j_{p}}}\Biggl[\ln {\frac{(m_{i}/m_{e})^{1/2}}{1+j_{p}}}\Biggr]^{-1}+1, 
\eqnum{A8}  \label{A8}
\end{equation}
and goes to $-\infty $ from the left side of the point $%
j_{p}=(m_{i}/m_{e})^{1/2}-1$, and to $+\infty $ from the right side.

It is readily seen that in the limit $j_p\to 0$, when charging is
predominantly collisional, and $a\ll \lambda_D$

\begin{equation}
\zeta \simeq \left( {\frac{m_{e}}{m_{i}}}\right) ^{1/2}{\frac{%
[(m_{i}/m_{e})^{1/2}-j_{p}]}{[\ln (m_{i}/m_{e})^{1/2}-j_{p}]}},  \eqnum{A9}
\label{A9}
\end{equation}
and

\begin{equation}
\left( {\frac{d\ln n_{d}}{d\ln n}}\right) _{q}\simeq 1-\left( {\frac{m_{i}}{%
m_{e}}}\right) ^{1/2}j_{p}.  \eqnum{A10}  \label{A10}
\end{equation}

\end{document}